\documentclass[sigconf,screen,authorversion,nonacm]{acmart}

\usepackage{fancyhdr}
\AtBeginDocument{%
    \addtolength{\footskip}{2.0\baselineskip}%
    \fancyfoot[L]{\textit{\textbf{Preprint --- final version forthcoming}}}%
}

\AtBeginDocument{%
  }

\usepackage{tabularx}
\usepackage[noend]{algorithm2e}
\RestyleAlgo{ruled}
\usepackage{enumitem}
\usepackage{pgf,tikz}
\usetikzlibrary{positioning}
\usepackage{tikz-network}
\usepackage{dsfont}
\usepackage{subcaption}
\usepackage{amsmath}
\usepackage{cleveref}
\usepackage{apxproof}
\crefname{algocf}{alg.}{algs.}
\Crefname{algocf}{Algorithm}{Algorithms}
\usepackage{booktabs}
\usepackage{amsthm}
\newtheoremrep{theorem}{Theorem}
\newtheorem{claim}{Claim}

\begin{document}

\title{Oh the Prices You'll See: Designing a Fair Exchange System to Mitigate Personalized Pricing}

\author{Aditya Karan}
\authornote{Both authors contributed equally to this research.}
\email{karan2@illinois.edu}
\affiliation{%
  \institution{University of Illinois at Urbana-Champaign}
  \country{USA}
}

\author{Naina Balepur}
\authornotemark[1]
\email{nainab2@illinois.edu}
\affiliation{%
  \institution{University of Illinois at Urbana-Champaign}
  \country{USA}
}

\author{Hari Sundaram}
\email{hs1@illinois.edu}
\affiliation{%
  \institution{University of Illinois at Urbana-Champaign}
  \country{USA}
  }

\renewcommand{\shortauthors}{Karan, Balepur, Sundaram}

\begin{abstract}

Many online marketplaces personalize prices based on consumer attributes. Since these prices are private, consumers may be unaware that they have spent more on a good than the lowest possible price, and cannot easily take action to pay less.
In this paper, we introduce a fairness-centered exchange system that takes advantage of personalized pricing, while still allowing consumers to individually benefit. Our system produces a matching of consumers to promote trading; the lower-paying consumer buys the good for the higher-paying consumer for some fee.
We explore various modeling choices and fairness targets to determine which schema will leave consumers best off, while also earning revenue for the system itself. 
We show that when consumers individually negotiate the transaction price, and our fairness objective is to minimize mean net cost, we are able to achieve the most fair outcomes. Conversely, when transaction prices are centrally set, consumers are often unwilling to transact. When price dispersion (or range) is high, the system can reduce the mean net cost to each individual by $66\%$, or the mean net cost to a group by $69\%$.
We find that a high dispersion of original prices is necessary for our system to be viable. Higher dispersion can actually lead to decreased net price paid by consumers, and act as a check against extreme personalization, increasing seller accountability. 
Our results provide theoretical evidence that such a system could improve fairness for consumers while sustaining itself financially.

\end{abstract}

\keywords{System Design, Fairness, Personalized Pricing}

\maketitle

\section{Introduction}

Suppose you and your friend are interested in buying the same pair of shoes online from the same website. You see the shoes listed for $\$60$, while your friend sees $\$50$ due to behavioral profiling~\cite{browsing23, hannak2014measuring}. 
If you knew about this price discrepancy, you could ask your friend to buy the shoes for you, paying them back and throwing in a few extra dollars so you can both buy the shoes at a lower net price.
Many such online marketplaces employ elements of personalized pricing \cite{consumer_experience}.
Differential pricing can vary with protected attributes \cite{Shiller2013First}, and fairness to consumers can suffer when personalized pricing is employed \cite{kallus2020fairness, moriarty2021online}.
While many consumers are aware of price personalization, it is difficult to take money-saving actions without knowing what other consumers are paying.

Fairer prices could be achieved either by sellers implementing fair pricing algorithms or via consumers' more active involvement in improving pricing outcomes.
On the seller's side, prior works examine fairer pricing algorithms~\cite{xu2022regulatory, grari2022fair, cohen2022price, banerjee2024fair, saxena2024unveiling}. These methods still attempt to build revenue maximizing, sometimes personalized, pricing algorithms while considering some notion of fairness. 
However, there is often not a way to enforce seller use of fair pricing algorithms. Thus, developing tools that promote consumer agency is key.
Using participatory design methods,
\citet{richards2016personalized} find that involving consumers in price-setting and negotiation leads to increased perceptions of fairness. 
Other prior work advises users to change their behaviors, and therefore receive better personalized prices~\cite{Kusner2017}. 
Another solution in online setting is coupon trading, which has been investigated as one means for consumers to get lower prices~\cite{kosmopoulou2016customer}.
However, none of these methods directly tackle achieve fairer pricing---allowing consumers to directly financially benefit without changing underlying behavior. 
A fairness-aware trading platform may be able to simultaneously reduce the impact of personalized pricing and improve fairness.

\begin{figure*}[t]
 \scalebox{1.1} {
      \centering
      \begin{tikzpicture}[scale = .7]
      \SetVertexStyle[MinSize = .5\DefaultUnit]
      \SetEdgeStyle[LineWidth = 1pt]

        \Text[x=0,y=-2.95]{(a) Market personalizes prices}
        \Text[x=0,y=-3.35]{for each agent for a good}

        \Vertex[x=0, y= 1.59, size = .1, style={color=white}, position = above, color = white]{pseudo}
        
        \Vertex[x=0, y=0, size = 2.25, shape = rectangle, label=Market, position = above, color = white]{market}
        
        \Vertex[x=0, y=0, label=$\$17$, RGB,color={211, 211, 211}]{n1}
        \Vertex[x=1, y=1,  label = $\$10$, RGB,color={211, 211, 211}]{n2}
        \Vertex[x=1, y=-1, label = $\$15$,  RGB,color={211, 211, 211}]{n4}

        \Edge[Direct, NotInBG = True](pseudo)(n1)
        \Edge[Direct, NotInBG = True](pseudo)(n2)
       
        \Edge[Direct, NotInBG = True](pseudo)(n4)

        \Text[x=5.5,y=-2.95]{(b) Agents choose to}
        \Text[x=5.5,y=-3.4]{interact with our system}

       \Vertex[x=5, y=0, size = 2.25, shape = rectangle, label=Market, position = above, color = white]{marketb}
        \Vertex[x=6.1, y=-.1, size = 2.2, shape = rectangle, label=Exchange System, position = below]{system}
        
        \Vertex[x=5, y=0, label=$\$17$, RGB,color={211, 211, 211}]{n1b}
        \Vertex[x=6, y=1,  label = $\$10$, RGB,color={211, 211, 211}]{n2b}
       
        \Vertex[x=6, y=-1, label = $\$15$,  RGB,color={211, 211, 211}]{n4b}

        \Text[x=10.95,y=-2.95]{(c) Our system gives a}
        \Text[x=10.95,y=-3.35]{matching; prices are set}

        \Vertex[x=10.5, y=0, size = 2.25, shape = rectangle, label=Market, position = above, color = white]{marketc}
        \Vertex[x=11.6, y=-.1, size = 2.2, shape = rectangle, label=Exchange System, position = below]{system}
        
        \Vertex[x=10.5, y=0, label=$\$17$, RGB,color={211, 211, 211}]{n1c}
        \Vertex[x=12.5, y=1,  label = $\$10$, RGB,color={211, 211, 211}]{n2c}
        \Vertex[x=12.5, y=-1, label = $\$15$,  RGB,color={211, 211, 211}]{n4c}

        \Edge[Direct, NotInBG = True, label = $\$12$](n4c)(n2c)
        \Edge[Direct, NotInBG = True, label = $\$16$](n1c)(n4c)

        \Text[x=16.5,y=-2.95]{(d) Mutually beneficial}
        \Text[x=16.5,y=-3.35]{transactions occur}

        \Vertex[x=16, y=0, size = 2.25, shape = rectangle, label=Market, position = above, color = white]{marketd}
        \Vertex[x=17.1, y=-.1, size = 2.2, shape = rectangle, label=Exchange System, position = below]{system}
        
        \Vertex[x=16, y=0, label=$\$17$, RGB,color={211, 211, 211}]{n1d}
        \Vertex[x=18, y=1,  label = $\$10$, RGB,color={211, 211, 211}]{n2d}
        \Vertex[x=18, y=-1, label = $\$15$,  RGB,color={211, 211, 211}]{n4d}

        \Edge[Direct, NotInBG = True,RGB,color = {76, 187, 23}, label = $\$12$](n4d)(n2d)
        \Edge[Direct, NotInBG = True,RGB, color = {170, 74, 68}, label = $\$16$](n1d)(n4d)

    \end{tikzpicture}
    }
    \caption{The relationship between the market (white square), the system (blue square), and consumers (gray circles). Consumers enter the market and are offered prices (a). They decide to participate in the exchange system (b). The system assigns a matching, and prices for these transactions are assigned or decided upon (c). The arrow direction represents the transfer of money. Of these matched transactions, only those that are mutually beneficial to both agents occur (d). This is determined by each agent's utility function.}\label{fig:tikz_intro} 
  \end{figure*}

In this work, we design a fairness-centered exchange system that allows consumers to trade to reduce the price they pay for a good online. 
As part of this system, we design matching and transaction-price setting procedures that aim to improve fairness.  
\Cref{fig:tikz_intro} depicts how the system, market, and consumers interact. 
To achieve our goal of designing an effective fairness-improving system,
we simulate various design choices.
With a given matching, we test two transaction price-setting procedures: centralized, where the system sets price, and decentralized, where consumers negotiate (\textit{RQ 1}). Our system is intended to improve fairness; we consider four fairness targets for each procedure. The four fairness targets cover
two objectives (the value measured: mean and variance of cost), and scopes (who it is measured for: individuals and groups). We investigate which fairness target is most feasible, or achievable, in our system (\textit{RQ 2}).
To ensure financial viability, we explore how the system's revenue is affected by the number of consumers, the cut the system takes from each transaction, and the dispersion (or range) of prices initially offered by the market (\textit{RQ 3}).
To conclude we examine a case study on an empirical pricing distribution modeled in prior work. 

We find that decentralized negotiations between consumers and minimizing the average net price are fairness maximizing. We are able to decrease the average price paid by $66\%$ when compared to the market without trades, under a highly dispersed pricing algorithm (producing a wide range of prices).   
Minimizing average net price paid per group leads to a $69\%$ decrease in the metric, compared to a $66\%$ decrease for average net price paid per individual. Minimizing standard deviation is unsuccessful, leading to increases in the metric. 
We find that as dispersion (or range) of prices increases, the exchange system acts as a more effective counterweight to unfair pricing. While in the high dispersion case we see a $66\%$ reduction in average price paid, the medium and low dispersion cases are a $15\%$ and $0\%$ reduction respectively.   
As we increase the cut taken by the system, the system earns more revenue; however, for a sufficiently high percent taken, system revenue falls as consumers find fees too high. 
Our contributions are as follow:

\textbf{Exchange system design:} We design an exchange system that could feasibly be implemented, financially sustain itself, and does not require the cooperation of the market. Prior work in this area has found personalized pricing in different markets \cite{hannak2014measuring, aznar2018airbnb, browsing23}, and some has proposed fair pricing algorithms \cite{aurangzeb2021fair, grari2022fair, xu2023doubly, xu2022regulatory, wang2016themis}. However, none to our knowledge has proposed an exchange system to address unfair personalized pricing. In this work we close that gap---exploring design choices and fairness targets that will lead to a financially self-sustaining, price-decreasing system.

\textbf{Opportunities for personalized pricing:} We find that even though personalized pricing can cause monetary loss to consumers, when properly taken advantage of, it can provide a money-saving opportunity. 
 While prior work notes personalized pricing can help some consumers by making certain goods more affordable to them \cite{NBERw23775}, we design a system that helps all consumers.
We show that, surprisingly, highly dispersed personalized pricing leads to lower costs for consumers \textit{and} higher revenue to the system.
This finding emphasizes the strengths of our exchange system design: we do not require sellers to cooperate by fairly pricing goods. In fact, when they don't, it can be beneficial to consumers and creates a measure of accountability to extreme, unfair personalization.

 \section{Related Work}

\textbf{Fairness and pricing:}
Various methods and metrics for fair machine learning have been proposed, in particular to deal with challenges of group fairness ~\cite{agarwal2018reductions, agarwal2019fair, Bera2019, somerstep2024algorithmic, small2024equalised}. In the pricing domain, unfair pricing and ad delivery caused by online behaviors have been found by several studies \cite{hannak2014measuring, aznar2018airbnb, browsing23, baumann2024fairness} and can result in structural inequity ~\cite{zhang2024structural}.

This work examines personalized pricing, which stands in contrast to flat or fixed pricing. While personalized pricing \textit{could} lead to consumer benefit \cite{bergemann2015limits}, it may also cause certain consumers to pay much more than others \cite{du2021fairness}, and lead to a perceived feeling of unfairness \cite{zuiderveen2017online}.
One solution is to redesign pricing algorithms for fairness. 
Seminal work on fair pricing \cite{heyman2008perceptions, rotemberg2011fair} examines what fair pricing is and how consumers may react to various pricing algorithms by firms. \citet{kallus2020fairness} examine how different markets and concerns (\textit{e.g.,} information asymmetry) 
inform which fairness criteria to consider. \citet{grari2022fair} study fair pricing under adverse selection, balancing actuarial risk with a demographic parity or equalized odds constraint. 
\citet{xu2022regulatory} explore how imposing restrictions on the degree of personalized pricing (\textit{i.e.,} the price of an object cannot vary more than $x\%$) can be customized to balance the needs of both buyers and sellers. 
The goal of fair pricing literature is to create fair algorithms that a seller can use while still achieving high profits.

While this approach\textit{ could }achieve lower prices for consumers \textit{if deployed}, this assumes the ability and willingness of a seller to do so. 
To mitigate this issue, we design a solution to unfair pricing without directly requiring cooperation from the seller. 
Some work has taken this approach---counterfactual fairness \cite{Kusner2017} could allow an individual to achieve better pricing without direct access to the underlying pricing system. However, this requires both an accurate causal model and toggles that individuals can act on \cite{10.1145/3442188.3445899}. On the collective action side, \citet{hardt2023algorithmic} quantified the effect of coordinated feature changes against an algorithmic system. In our work, collaboration does not aim to change the direct outcome of an algorithmic system, but rather to adjust final outcomes for individuals.  
Our work does not require a causal model or online behavior change for individuals to benefit. Rather, we focus on consumers who are incentivized by monetary reward to share information. We specifically are interested in cases where the goods are identical and purchased on online marketplaces, where potentially unfair differential pricing can occur without users' knowledge \cite{Lippert2023, miller2014we}.

\textbf{Federated learning:}
In federated learning, data is indirectly ``shared'' via model updates and aggregated (\textit{e.g.,} FedAvg ~\cite{pmlr-v54-mcmahan17a}). 
There is a very large existing body of work investigating fairness and trying to achieve it, on both group \cite{lyu2020collaborative, ezzeldin2023fairfed, zeng2021improving, yu2020fairness, huang2020fairness, li2021ditto, du2021fairness, li2019fair} and individual \cite{LI2023103336, Yue_2023} levels. In the fair federated learning framework, it is already assumed that individual entities want to participate, as the outcomes are clear. Participation can improve both accuracy in the underlying models as well as some notion of fairness.

In our work, we focus on whether we can incentivize rational agents who act purely selfishly to achieve some notion of fairness. \citet{donahue2021model} extend the federated learning context to consider whether agents should participate in a shared model or rely on only their local information. However, this work differs from ours as our objective is not one of model performance. They further investigate \cite{kleinberg2023sharing} egalitarian and proportional fairness in the context of these model-sharing games. \citet{salehi2012} develop a model-sharing architecture for agents' mental models but do not explicitly consider fairness. 

\textbf{Marketplace mechanisms:} \citet{shapley1971assignment} and \citet{roth1992two} both examine the assignment game --- a two-sided matching with money.
In this work, we are interested in situations where everyone has access to the same good at different prices, rather than valuing goods at different prices. \citet{jagadeesan2021matching} look at how some markets give rise to universal pricing, while others employ personalized pricing. 
\citet{babaioff2021making} and \citet{branzei2024learning} design auctions to limit gains from post auction dealings, similar to those we introduce here.

\section{Problem Statement} \label{sec:problem}

Consider a marketplace $\mathcal{M}$ that offers one type of good $g$ to consumers $V$, where $N = |V|$. Each consumer $v \in V$ requires exactly one unit of good $g$; the supply of $g$ is finite but sufficient to satisfy $V$. Each consumer $v \in V$ has a vector of attributes $r_v$, a resource constraint $k$, and a private utility function $f_v$. The consumers in $V$ can be partitioned into non-overlapping groups based on $r_v$. 

Pricing algorithm $\mathcal{A}$ takes a vector of properties of the consumer $r_v$ as input, and outputs a personalized price $p_v$ for consumer $v$. Resulting prices may have varying degrees of personalization---we measure this using the \textit{dispersion} of prices, $\delta$, our measure of pricing spread.
In this work, we design an exchange system $\mathcal{S}$ that takes advantage of personalized pricing so consumers profit. System $\mathcal{S}$ pairs up consumers via matching process $\mathcal{P}$, which outputs a set of pairwise consumer interactions $\mathcal{J}$. Let $j_{uv} \in \mathcal{J}$ be the (directed) interaction parameterized by consumers $u$ and $v$, where $p_u > p_v$. In an interaction $j_{uv}$, consumer $u$ will pay some $m$ dollars to consumer $v$ for good $g$. The system $\mathcal{S}$ then takes some fraction $\gamma$ of this $m$ to sustain the trading ecosystem, so the payment $v$ will receive for this transaction is $(1-\gamma)m$. The interactions in $\mathcal{J}$ are proposed to consumers by the matching process $\mathcal{P}$, but all interactions need not occur. An interaction is executed if both consumers $u$ and $v$ benefit according to their utility functions $f_u$ and $f_v$.

We frame the matching process $\mathcal{P}$ as a network problem. Our consumers exist in a directed network $G = (V, E)$ where $(u,v) \in E $ if $p_u > p_v$. Interactions $\mathcal{J}$ are matched directed edges; each consumer $v$ has at most $k$ matched incoming edges. The outgoing edge count is capped at 1, since each consumer can only buy the good $g$ once. This resource constraint $k$ represents how much time a consumer is willing to spend on this system.
If an interaction $j_{uv}$ is executed, the edge $(u,v)$ has been transacted on. We call $u$ the buyer and $v$ intermediary; one consumer might be matched as both an intermediary and a buyer on different transactions.

In this paper, we examine which design of exchange system $\mathcal{S}$ will maximally improve our fairness targets $\mathcal{F}$ on market $\mathcal{M}$, and under what circumstances it is financially feasible to maintain.
Specifically we ask:
\begin{description}
    \item[RQ 1:] Given a fairness target $\mathcal{F}$, which of the following methods to set transaction price $m$ will maximize fairness: centralized setting or individual negotiation?
    
    \item[RQ 2:] In the context of our exchange system $\mathcal{S}$, are some fairness considerations (\textit{i.e.,} definitions of $\mathcal{F}$) more feasible to achieve than others? In particular, how does varying $\mathcal{F}$ in scope and objective affect feasibility?
    
    \item [RQ 3:] How does revenue to the exchange system $\mathcal{S}$ change with the number of consumers $N$, cut $\gamma$ taken by the system, and dispersion $\delta$ of the pricing algorithm $\mathcal{A}$?

\end{description}

\section{Fairness}\label{sec:fairness} %
In \textbf{RQ 2} we ask whether some fairness criteria $\mathcal{F}$ are more feasible than others to achieve. Inspired by \citet{kallus2020fairness}, we develop our notions of fairness to be suited to this setting. We started by considering what the ``ideal'' scenario would be for all consumers. We determined the ideal fair outcome would result in all consumers paying the \textit{same, lowest} price for the good. Hence, we seek to minimize both average price and the standard deviation in prices paid by consumers. However, these ideal outcomes might not be possible in our exchange system. Thus, we demonstrate lower bounds for mean (Theorem 1) and standard deviation (Claim 1) which can be found in \Cref{subsec:theory}. We measure the ``feasibility'' of our fairness metrics by considering how close each mean and standard deviation come to the ideals.

Thus, we vary objective and scope for our fairness definitions, \textit{i.e.,} \textit{what} we measure, and \textit{who} we measure it for. 
All objectives take as input the net cost incurred by each consumer from participating in system $\mathcal{S}$, assuming that each consumer purchases exactly one unit of good $g$. For example, if consumer $u$ is offered price $p_u$ by the market, but pays $m$ dollars to consumer $v$ to buy good $g$ for price $p_v$, then consumer $u$'s net cost is $m$. Consumer $v$'s profit from selling the ticket is  $(m(1-\gamma) - p_v)$ and therefore their net cost for their own ticket is $p_v - (m(1-\gamma) - p_v)$. As a reminder, consumers in our system can engage in up to $k$ transactions as intermediary.
We denote $\omega_u$ as the net cost to consumer $u$, where $\omega_u < 0$ implies monetary gain to $u$. We define two objectives (mean and standard deviation) and two scopes (individual and group-level). We define a group as a collection of individuals who share a demographic attribute (in our case, $r_v$). We call this set of groups $\mathcal{G}$. This gives us four different fairness measures; we present them in \Cref{tab:fairness}.

\begin{table}
\centering
    \caption{Our fairness outcome measures. All measures rely on $\omega_u$, the net cost to consumers $u$. $\mu_g$ represents the average net cost to consumers $u$ in group $g$. $|\mathcal{G}|$ is the number of groups.}
    \label{tab:fairness}
    \begin{tabularx}{.9\linewidth}{@{}r c c@{}}
    & \textsc{Individual} & \textsc{Group}\\ 
    \midrule
        \textsc{Mean} &\normalsize $\mu_I = \frac{\sum_{u \in V} \omega_u}{N} $ & \normalsize$\mu_\mathcal{G} = \frac{\sum_{g \in \mathcal{G}} \mu_g}{|\mathcal{G}|}$\\\\
        \textsc{S.D.} & \small$\sigma_I = \sqrt{\frac{\sum_{u \in V}(\omega_u - \mu_I)^2}{N}}$ &\small $\sigma_\mathcal{G} = \sqrt{\frac{\sum_{g \in\mathcal{G}}(\mu_g - \mu_\mathcal{G})^2}{|\mathcal{G}|}}$ 
\end{tabularx}
\end{table}

We seek to minimize these values. In minimizing individual mean, we want all consumers to minimize the price they're paying (some may even earn money from $\mathcal{S}$).
Minimizing individual standard deviation captures that we want our system's benefit to be distributed equally among consumers; no single consumer should benefit while others do not. In the group fairness cases, we desire similar outcomes, but distill each group by taking the mean over individuals in the group. We recognize that these definitions of fairness are neither standard nor exhaustive. We considered more standard fairness metrics such as demographic parity, predicted parity, and equalized odds \cite{agarwal2018reductions,agarwal2019fair, Bera2019, Zemel2013, Zafar2017, Kusner2017, Kleinberg2017, Dwork2012}. 
However, we decided that fairness would best be captured in this application by metrics designed from our ideal scenario (where all individuals get the lowest price). 
We vary our fairness target over these definitions to test \textbf{RQ 2}.

\section{The Model}\label{sec:model}

\subsection{Model overview}
We simulate our proposed exchange system via rational agents. We begin with consumers $V$ who desire exactly one unit of good $g$ on market $\mathcal{M}$.
Each consumer $v \in V$ is initialized with attribute vector $r_v$, utility function $f_v$, and price $p_v$ for good $g$ according to pricing algorithm $\mathcal{A}$. 

The matching process $\mathcal{P}$ outputs a set of interactions $\mathcal{J}$ between consumers. The transaction price $m$ of each interaction is then set in either a centralized or decentralized fashion (\textit{i.e.,} either set by the system or via consumer negotiations). Given a matched edge $(u,v)$, buyer $u$ and intermediary $v$ exchange money for good $g$ if $u$ and $v$ both have positive utility for the transaction and are within their resource constraints. Once all  consumers make transaction decisions, any consumer $x$ who has not yet bought good $g$ will 
do so at price $p_x$, the original offered price.
\Cref{alg:overview} below references sub-routines 
$getPrice$, $solveMatching$ and $computeM$, which we detail in this section, along with utility function $f_u$ for consumer $u$.

\subsection{Consumer attributes} \label{subsec:agents}

\textbf{Consumer properties:} We assign each consumer $v$ a vector of consumer properties $r_v$.
We implement $r_v$ as a scalar, but this can be trivially extended as a vector. 
We consider this consumer property to represent some demographic feature that is used by the pricing algorithm $\mathcal{A}$ to assign price $p_v$.

\textbf{Resource constraint:} Each consumer $v \in V$ has the same resource constraint $k$---the number of interactions for which consumer $v$ can serve as intermediary. All consumers serve as buyer for one interaction.

\textbf{Utility function:} Each consumer $u$ has a utility function $f_u$. The form of $f$ is the same for each consumer. Recall that in a given interaction $j_{uv}$, consumer $u$ is the buyer while consumer $v$ is the intermediary. The utility function for $u$ can be written as 
$f_u(j_{uv}) = p_u - m - \epsilon_{u_j}$ and for $v$ as $f_v(j_{uv}) = m(1-\gamma) - p_v - \epsilon_{v_j}$ where the $(1-\gamma)$ term accounts for the system receiving $\gamma$ proportion of the transaction amount $m$.
Terms $\epsilon_{u_j}$ and $\epsilon_{v_j}$ are the disutility to $u$ and $v$ from spending time on this interaction. Each consumer $u$ is assigned a truncated Normal distribution $\mathcal{E}_u$, and $\epsilon_{u_j}$ is drawn \textit{i.i.d.} from $\mathcal{E}_u$.
\begin{algorithm}[t]
\caption{Model overview}\label{alg:overview}
\KwData{$V$ the set of consumers}
\KwResult{$G$ the network after trades, including transaction prices}
\tcp{PRICING ALGORITHM}
\For{$u \in V$}{
  $p_u \gets$ $getPrice(u)$; \quad \tcp{market assigns personalized price to consumer}
}
\tcp{MATCHING PROCESS}
$D$ an empty map from $V \to V$ \;
$G \gets (V,E)$ such that directed edge $(u,v) \in E$ exists iff $p_u > p_v$ \;
\For{$u \in V$}{
  $D(u) \gets solveMatching(u,G,k)$; \quad \tcp{produces matching of consumers}  
}
\tcp{EXCHANGE PROCESS}
\For{$u \in V$}{

$v = D(u)$ \;
$m \gets computeM(j_{uv})$ ; \quad \tcp{gets $m$ in a (de)centralized manner}  
\If{$f_u(j_{uv})>0$ and $f_v(j_{uv})>0$}{
        consumer $u$ pays $m$ dollars to $v$ to get good $g$ at price $p_v$ \;
  }

\Else{
consumer $u$ pays $p_u$ to the market $\mathcal{M}$ for good $g$
}
}
\Return{$G$}
\end{algorithm} 
\subsection{Pricing algorithm $\mathcal{A}$ $(getPrice)$}\label{subsec:pricing}
Pricing algorithm $\mathcal{A}$ determines the price for good $g$ offered to each consumer in $\mathcal{M}$ ($getPrice$). 
To capture a wide range of pricing algorithm behavior, we define a notion of \textit{dispersion} ($\delta$) which represents the spread of prices outputted by algorithm $\mathcal{A}$ when prices are normalized to $[0, 100]$. Mathematically, it is the range within which a large proportion of possible prices fall. We define a pricing algorithm $\mathcal{A}_\delta$ which assigns prices with dispersion $\delta$, and has the following properties. 1) The range of feasible prices is in $(0,100]$. 2) $\mathcal{A}_\delta$ is biased based on some immutable attribute $r_v$ of the consumer $v$. We partition the set of consumers $V$ into non-overlapping groups based on attribute $r_v$. We call this set of groups $\mathcal{G}$. 3) Consumer $v$'s price $p_v$ is drawn from a distribution $\mathcal{D}_{\mathcal{G}_v}$, where ${\mathcal{G}_v}$ is $v$'s group. All consumers from the same group have their price drawn from the same distribution.
4) Dispersion $\delta$ represents the $2.25\sigma$ range of possible prices (\textit{i.e.,} 
    $\max_{g, h \in \mathcal{G}}  ((\mu_g - 2.25\sigma_{g}) - (\mu_{h} + 2.25\sigma_h)))$, where $\mu_g$ represents the average net cost to consumers $u$ in group $g$.
    
In addition to our simulated prices we also examine an empirical pricing algorithm; we use a pricing model presented in prior work \cite{browsing23}, which investigates airline ticket pricing. 
We assume that prices shared by consumers are collected by an automated system (\textit{e.g.,} browser extension) to ensure that prices available to the algorithm are the same prices available to the consumer. 

\subsection{Exchange system $\mathcal{S}$ $(solveMatching \textmd{ and } computeM)$ }\label{subsec: model_system}

In \textbf{RQ 1} we ask whether the transaction price $m$ should be set centrally or determined by individual negotiations. We experiment with two methodologies for setting the transaction price $m$:

\textbf{Centralized:} We solve for the optimally fair consumer matching and $m$-value setting (maximizing fairness criteria $\mathcal{F}$). The system is not aware of private utility functions of individuals, so recommended matchings at optimal $m$ values may not actually transact. This optimization outputs a set of interactions $\mathcal{J}$ such that each consumer has at most one outgoing interaction (\textit{i.e.,} as buyer), and at most $k$-many incoming interactions (\textit{i.e.,} as intermediary). The transaction prices $m_{uv}$ for all $j_{uv} \in \mathcal{J}$ are simultaneously set centrally to maximize $\mathcal{F}$. Each consumer can either accept the transaction price $m$ or refuse.
 
 Below is the linear program using $\mu_I$, the fairness objective that represents individual mean net cost paid, as our example objective function. The other objectives can be found in \Cref{tab:fairness}. The goal of this program is to output interactions $\mathcal{J}$ along with transaction prices $m_{uv}$ for all $j_{uv} \in \mathcal{J}$. If no $j_{uv}$ exists in $\mathcal{J}$ for a consumer $u$, then $u$ will pay the original price it was assigned for good $g$, which is $p_u$. 
Due to the presence of the binary variable $x_{uv}$, this is a mixed integer program, which is at least NP-hard \cite{arora2009computational}. Minimizing $\mu_{I}$ and $\mu_{\mathcal{G}}$ is a linear objective, while minimizing $\sigma_{I}$ or $\sigma_{\mathcal{G}}$ results in a Mixed Integer Quadratic Program, also NP-hard ~\cite{pia2017mixed}. 
In our analysis for the linear objective, we solve to completion. For the quadratic objective, we return the best result given by the solver after a pre-specified amount of time (60 seconds). We use Gurobi \cite{gurobi} as our solver. 
    \begin{align*}
       & \text{minimize }    \quad \frac{\sum_{u \in V} \omega_u}{|V|} \\ 
        & \text{s.t. } \quad p_v/(1-\gamma) \leq m_{uv} \leq p_u \quad \forall j_{uv} \in \mathcal{J} \\
        & \quad x_{uv} \in \{0, 1\}  \quad \forall j_{uv} \in \mathcal{J} \quad \text{// $u$ buys from $v$ }  \\
          & \quad \sum_{v} x_{uv} \leq 1 \quad \forall u \in V \quad \text{// a consumer can buy once}\\
         & \quad \sum_{u} x_{uv} \leq k \quad \forall v \in V \quad \text{// and act as intermediary $k$ times}\\ 
         & \quad \omega_{u} = \sum_{v\in V}\underbrace{x_{uv}m_{uv}}_{\text{buyer's new price if traded}} + \sum_{v\in V}\underbrace{(1 - x_{uv})p_u}_{\text{buyer's new price if not traded}} \\
        & \quad \quad -  \sum_{v\in V} \underbrace{x_{vu}(m_{uv}(1-\gamma) - p_v)}_{\text{intermediary's profit if traded}}  \quad \forall u \in V 
     \label{eq:opt_centr}
    \end{align*}

\textbf{Decentralized (individual negotiation):} Here, we keep the matching $\mathcal{J}$ given by the centralized process, but allow consumers $u$ and $v$ to negotiate individually for the transaction prices $m_{uv}$.
Rather than simulate bargaining between the consumers, we make an assumption regarding the settled transaction price.
In our implementation, we set each $m_{uv}$ to the Nash bargaining solution, which maximizes the product of the net prices paid ~\cite{osborne1990bargaining}. We choose this value because it is a likely outcome after individual negotiations. Other values could be used as well, such as the mean of prices. 

We test four fairness definitions, $\mathcal{F}$, and two price setting processes for $m$; in total this gives eight methodologies. For convenience we denote each methodology as a fairness metric, $X$ indexed by a process, $Y$, \textit{i.e.,} $X^Y$ for $X \in \{\mu_{I}, \mu_{\mathcal{G}}, \sigma_{I}, \sigma_{\mathcal{G}} \}$ and $Y \in$ \{C, D\}. For example, $\mu_{I}^{C}$ refers to optimizing $\mu_{I}$ via the centralized methodology. We also notate $\mu^Y$ to refer to $\mu_{I}^Y, \mu_{\mathcal{G}}^{Y}$ methods and 
$\sigma^Y$ to refer to $\sigma_{I}^{Y}, \sigma_{\mathcal{G}}^Y$ methods. 

\subsection{Theoretical claims}\label{subsec:theory}

\begin{theorem}
\label{theorem:mean}
For $\epsilon_u = 0$ $\forall u \in V$, the mean net cost over all consumers after trading is bounded below by $p_{min} (\frac{1 - \frac{|B|\gamma}{N}}{1 - \gamma})$, where $B$ is the set of consumers that will never engage as a buyer in the system, and $p_{min}$ is the minimum price assigned to any consumer.
\end{theorem}
\begin{proof}

Without loss of generality, let consumer $u$ buy a ticket from consumer $v$ for a cost of $m_{uv}$. Let us define gain to a consumer from trade as the cost improvement the consumer experiences as a result of a given transaction. For a fixed $\gamma$ (the proportion of the trade cost taken by the system),
the gain to the intermediary from this interaction is $m_{uv}(1-\gamma) - p_v$ while the gain to the buyer is $p_u - m_{uv}$.
Then, the sum of the gain to consumers $u$ and $v$ from this transaction is $(p_u-m_{uv}) + (m_{uv}(1 -\gamma) - p_v) = p_u-p_v-\gamma m_{uv}$. 

In order to minimize mean net cost, we must maximize gains from trade across all consumers for all possible transactions.
To maximize gain from a single trade between consumers $u$ and $v$, $m_{uv}$ must be minimized. 
The minimum value for $m_{uv}$ is the transaction price $m_{uv} = \frac{p_v}{1-\gamma}$. When $\epsilon_u = 0$, at this price the intermediary $v$ experiences $0$ gain; a buyer will accept this price if $p_u \geq \frac{p_v}{1-\gamma}$. Denote $A$ as the set of consumer pairs $(u, v)$ that have traded and $B$ as the set of consumers $u \in V$ that have not acted as a buyer. We note that $|A| + |B| = N$ as a consumer can only be a buyer once. 
We can then write the net price paid as shown below, since  $p_{min}$ is the lowest price possible. 

$$\sum_{(u, v) \in A} \frac{p_v}{1-\gamma} + \sum_{u \in B} p_u \geq |A|\frac{p_{min}}{1-\gamma} + |B| p_{min}$$
$$=\frac{(|A| + |B|)p_{min} - |B| \gamma p_{min}}{1 - \gamma} = p_{min}\left( \frac{N - |B| \gamma}{1 - \gamma}\right) $$

Dividing by $N$ we are left with the following lower bound:
$$p_{min} \left(\frac{1 - \frac{|B|\gamma}{N}}{1 - \gamma}\right)$$ 
\end{proof}

\begin{claim}
\label{theorem:std}
When $\gamma = 0$ and $k = |V| - 1$, the minimum standard deviation in net cost is $0$. 
\end{claim}
\begin{proof}
When $\gamma = 0$, minimizing standard deviation in net price requires that all consumers trade with the consumer who got the lowest price at $m = p_{min}$. Then, all buyers will pay $p_{min}$, while the intermediary consumer earns nothing, and still pays $p_{min}$ for their good. The net cost to all consumers is $p_{min}$ and thus the standard deviation is $0$. 
\end{proof}

\begin{claim}
\label{theorem:IR}
This system satisfies Individual-Rationality (IR).  
\end{claim}

\begin{proof}
    Buyers transact at price $m$ if 
    $f_u(j_{uv}) = p_u - m - \epsilon_{u_i} \geq 0$.
    If both consumers have positive utility they will trade and benefit. Otherwise, one or both will reject and both will pay their original price. Hence, they are no worse off from participating.
\end{proof}

 \begin{figure*}[h]
    \centering
    \begin{subfigure}{.45\linewidth}
        \centering
        \includegraphics[width=\linewidth]{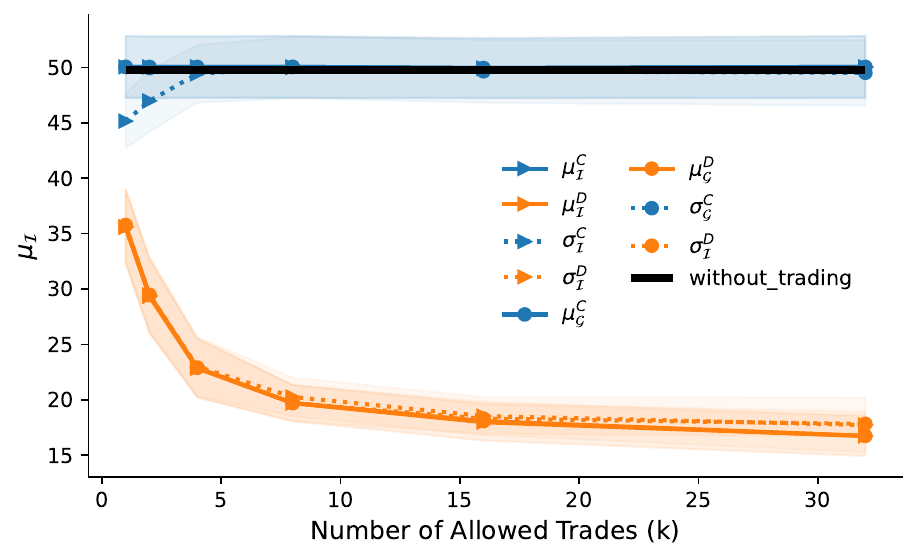}
        \caption{Individual mean net cost, $\mu_I$}
        \label{fig:rq1a}
    \end{subfigure}
    \begin{subfigure}{.45\linewidth}
        \centering
        \includegraphics[width=\linewidth]{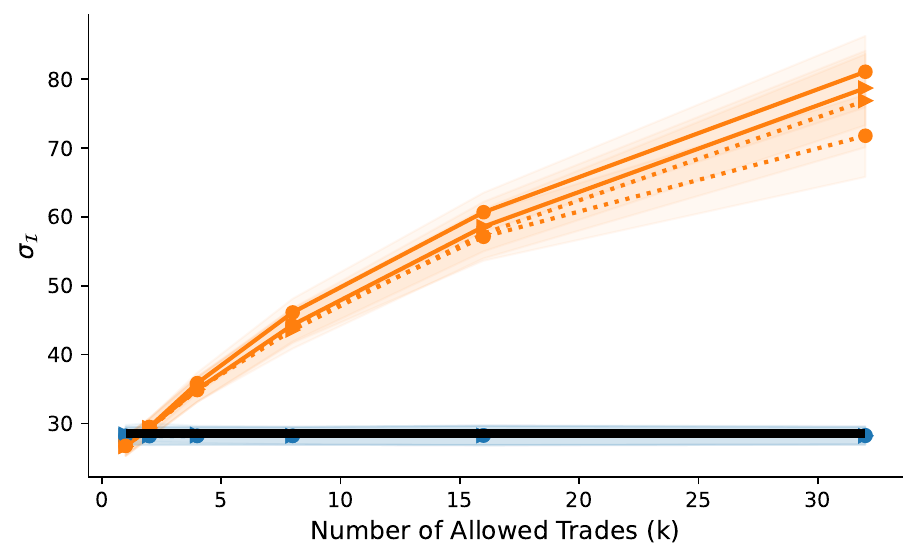}
        \caption{Individual std. in net cost, $\sigma_I$}
        \label{fig:rq1b}
    \end{subfigure}
    \begin{subfigure}{.45\linewidth}
        \centering
        \includegraphics[width=\linewidth]{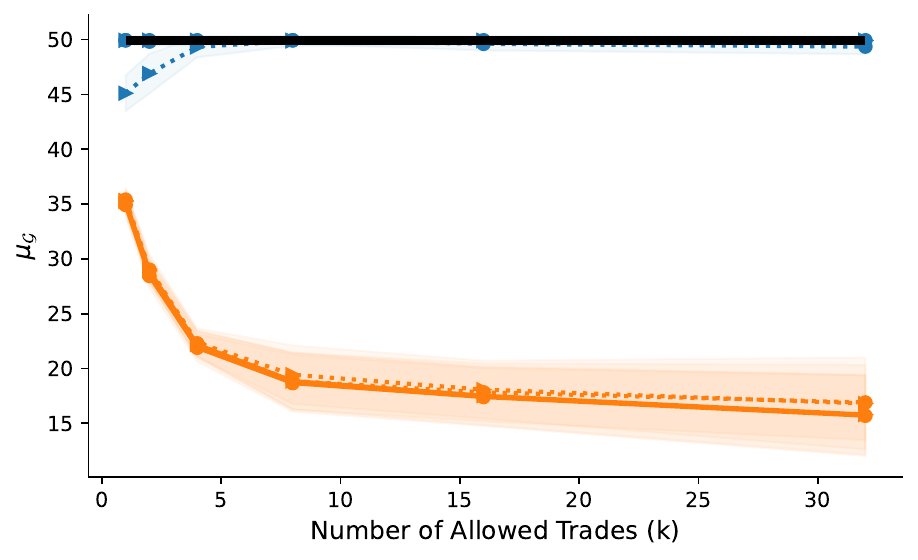}
        \caption{Group mean net cost, $\mu_{\mathcal{G}}$}
        \label{fig:rq1c}
    \end{subfigure}
    \begin{subfigure}{.45\linewidth}
        \centering
        \includegraphics[width=\linewidth]{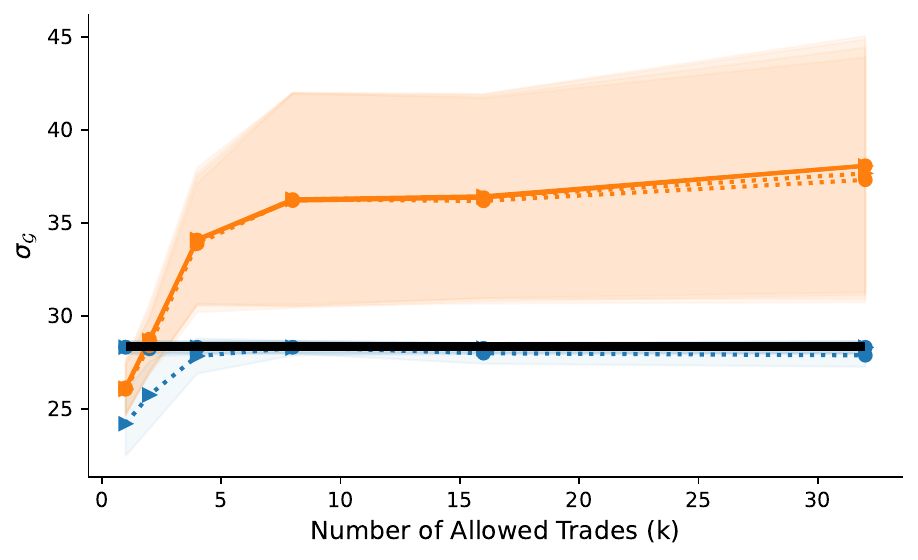}
        \caption{Group std. in net cost, $\sigma_{\mathcal{G}}$}
        \label{fig:rq1d}
    \end{subfigure}
    \caption{Realization of four fairness definitions (a-d) under eight different optimizations (eight curves representing four fairness targets, two price-setting procedures). We set $N= 100, \gamma = 0.4$ and use pricing algorithm $\mathcal{A}_{0.95}$. Centralized methods are blue; decentralized orange. Variance bands show one standard deviation.\textbf{ $\mu_{I}$ and $\mu_{\mathcal{G}}$ can be effectively lowered by decentralized methods} targeting their respective metrics. $\sigma_{I}$ and $\sigma_{\mathcal{G}}$ are less feasible for all metrics; in particular $\sigma_{I}$ for decentralized methods increases with $k$.}
    \label{fig:rq1_gamma0}
\end{figure*}
\section{Results}

We implement our model $\mathcal{S}$ with 100 consumers ($N = 100$). We sample consumer disutility from
$\mathcal{E}_u \sim \text{truncated } N(\mu_u, 1)$ where $\mu_u\sim U(0,2)$.
We set $|\mathcal{G}| = 5$ where every consumer has an equal probability of being assigned to each group. We choose dispersion values $\delta = \{0.05$, $0.25,$ $0.50$, $0.75$, $0.95\}$ which maps to five pricing algorithms: $ \mathcal{A}_{0.05}$, $\mathcal{A}_{0.25}$, $ \mathcal{A}_{0.50}$, $ \mathcal{A}_{0.75}$, $ \mathcal{A}_{0.95}$; the exact mapping can be found in \Cref{sec:pricing}. In our construction, we change the distribution of prices while ensuring that mean price across pricing algorithms is roughly equal ($\$50$) which allows us to analyze the impact of dispersion directly. All pricing algorithms are constructed by a set of Normal distributions so prices are within $(0, 100]$. For example, $\mathcal{A}_{0.95}$ involves five Normal distributions with means $\{10, 30, 50, 70, 90\}$ respectively, all with standard deviation $\frac{10}{3}$. 

In our analysis for the linear objective, we solve to completion. For the quadratic objective, we return the best result given by the solver after a pre-specified amount of time (60 seconds). We use Gurobi \cite{gurobi} as our solver. 
In regards to optimization feasiblity---we find the duality gaps for $\mu_{\mathcal{I}}$ and $\mu_{\mathcal{G}}$ are $0$. For $\sigma_{\mathcal{I}}$, the duality gap is $67\%$, while for $\sigma_{\mathcal{G}}$ it is $0.02\%$. This suggests that $\mu_{\mathcal{I}}$ and $\mu_{\mathcal{G}}$ are more feasible to optimize, while $\sigma_{\mathcal{I}}$ is more difficult. We discuss this feasibility further in \Cref{sec:result_rq2}.

\subsection{RQ 1: Centralized versus decentralized $m$-setting}

In \textbf{RQ 1} we ask: Given a fairness target $\mathcal{F}$ (\Cref{tab:fairness}), which of the following methods to set transaction price $m$ will maximize fairness: individual negotiation or central allocation? In \Cref{sec:model} we described in detail two methods to set transaction price $m$ for an interaction $j_{uv}$, where $u$ buys from intermediary $v$. In brief, the centralized setting solves for
optimal edges and prices $m$, from which consumers can choose to accept or reject the given price. The decentralized option uses these same solved edges, but instead allows paired consumers to negotiate for the transaction price $m$.  
In our implementation, $m$ values for these interactions are chosen according to the Nash bargaining solution.

To answer \textbf{RQ 1}, we simulate transactions on our system $\mathcal{S}$. We test the centralized and decentralized procedures for setting $m$, four fairness targets $\mathcal{F}$ (\textit{i.e., } eight cases), six values for $k$ (the number of interactions for which a consumer can serve as intermediary), with $\gamma = 0.4$ (the cut the system takes from the interaction). Here we present $\mathcal{A}_{0.95}$ as the pricing algorithm (prices are highly dispersed with $\delta = 0.95$). Each simulation is run 100 times; we present the average of these, and bands that show one standard deviation. 
In \Cref{fig:rq1_gamma0}, we plot all eight cases against four fairness targets. 

Notably, the centralized procedures vary minimally and have little impact on fairness targets (\Cref{fig:rq1_gamma0}). Under this procedure, each consumer is required to either accept or reject the price that maximizes fairness globally. As described in Theorem 1, to minimize mean net cost, a consumer acting as intermediary will be asked to engage in the exchange without benefiting monetarily. Because each consumer experiences disutility $(\epsilon)$ that is unknown to the system, very few intermediaries engage in exchanges at centrally-set prices. The same is true for other fairness targets.

We contrast this with decentralized procedures where pairings are set, but consumers negotiate the prices. These negotiated prices are more likely to provide utility to both consumers, and hence more trades occur. When optimizing for $\mu_{I}$ (individual mean; shown in \Cref{fig:rq1a}) and $\mu_{\mathcal{G}}$ (group mean; shown in \Cref{fig:rq1c}) this results in the decentralized procedure lowering the net cost to an average of $\$17$ (\Cref{fig:rq1a})--- nearly a $66\%$ reduction in cost, compared to the average assigned price of $\$50$. However, even when edges are selected to minimize the standard deviation at the individual ($\sigma_{I}$; \Cref{fig:rq1b}) or group ($\sigma_{\mathcal{G}}$; \Cref{fig:rq1d}) level, the decentralized procedure can lead to increased price variation as $k$ increases, because consumers selfishly negotiate. As $k$ increases, consumers with lower prices can complete more transactions (benefiting more monetarily), while those with high prices can only benefit by buying once.

\subsection{RQ 2: Feasibility of our fairness measures}
\label{sec:result_rq2}

\begin{table}[h]
\caption{Our ``feasibility'' measure for each fairness metric when $k = 32, \gamma = 0.4$. We examine whether optimizing for each metric actually improves said metric. $\mu^D$ methodologies are feasible while $\sigma^D$ methodologies are not.}
\begin{tabularx}{.9\linewidth}{r|c|c|r}
\centering
Metric & Pre-exchange & Post-exchange &  \% \textbf{Change} \\ \hline
\multicolumn{1}{l|}{$\mu_{I}$}       & 49.8               & 16.7                 & \textbf{-66\%}     \\
\multicolumn{1}{l|}{$\sigma_{I}$}        & 28.5             & 76.9                 & \textbf{170\%}     \\
\multicolumn{1}{l|}{$\mu_{\mathcal{G}}$} & 49.9             & 15.7                 & \textbf{-69\%}    \\
\multicolumn{1}{l|}{$\sigma_{\mathcal{G}}$}  & 28.3           & 37.3                 & \textbf{32\%}     
\end{tabularx}
\label{tab:effectiveness}
\end{table}

In \textbf{RQ 2} we ask whether some fairness conditions (\textit{i.e.,} definitions of $\mathcal{F}$, described in \Cref{sec:fairness}) are more feasible to achieve than others. In particular, how does varying $\mathcal{F}$ in scope (\textit{i.e.,} individual vs. group) and objective (mean vs. variance) affect the ability of the system to achieve that fairness outcome?
To answer this question we run simulations as described previously.

In \Cref{fig:rq1_gamma0} we show how optimizing for each fairness target $\mathcal{F}$ affects the measure of each outcome. Here we will focus on the decentralized scenario, as centralized price-setting results in minimal trades. We see that when specifically optimizing for $\mu_{I}$ and $\mu_{\mathcal{G}}$, decentralized price-setting is able to effectively reduce the average price paid. On the other hand, achieving low standard deviation (\Cref{fig:rq1b,fig:rq1d}) is more difficult. We can investigate how optimizing for each fairness metric compares to the ``ideal'' scenario. In Table \ref{tab:effectiveness} we show numerically that $\mu$-optimizing procedures are more feasible under decentralized settings, while $\sigma$-optimizing are less so. During decentralized negotiations, as $k$ (the number of times a consumer can act as intermediary) increases, intermediaries are able to benefit multiple times while buyers only benefit once. This increases the gap, even when choosing $\sigma$-optimizing pairings.

\subsection{RQ 3: Revenue to the exchange system $\mathcal{S}$}
\begin{figure}[h]
    \includegraphics[width=.95\linewidth]{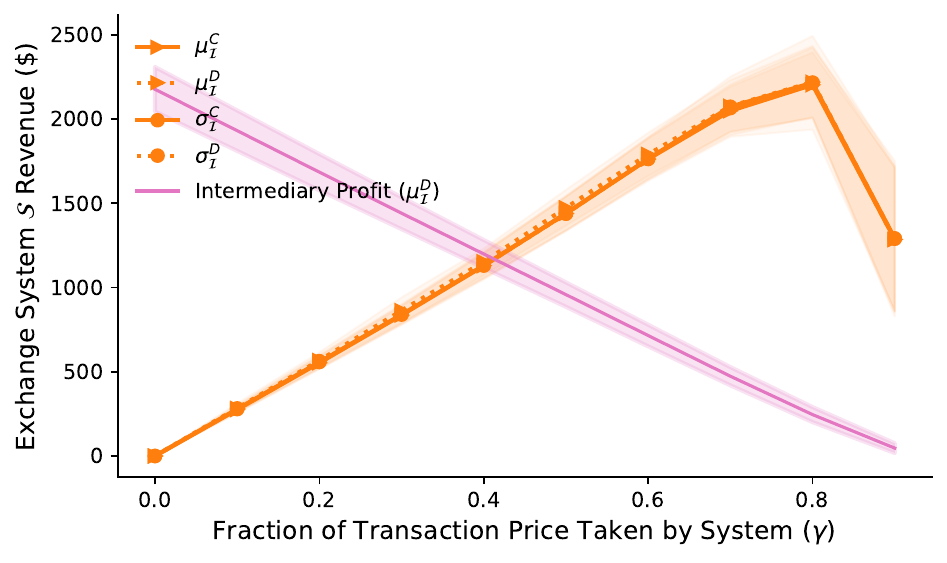}
    \caption{
       System revenue for four methodologies and intermediaries' profits for $\mu_{\mathcal{I}}^D$. System revenue changes with respect to $\gamma$ under high dispersion $\delta = 0.95$ and $k = 16$. \textbf{The system at most earns $\approx \$2200$ in revenue, representing a large cut to the original seller's revenue without trading.}
    For sufficiently high $\gamma$, system revenue falls as consumers find fees too high. Intermediary profits decline as a function of $\gamma$.
    } \label{fig:rq3}
\end{figure}
    \begin{figure}[h]
    \centering
    \begin{subfigure}{0.95\linewidth}
        \centering
        \includegraphics[width=\linewidth]{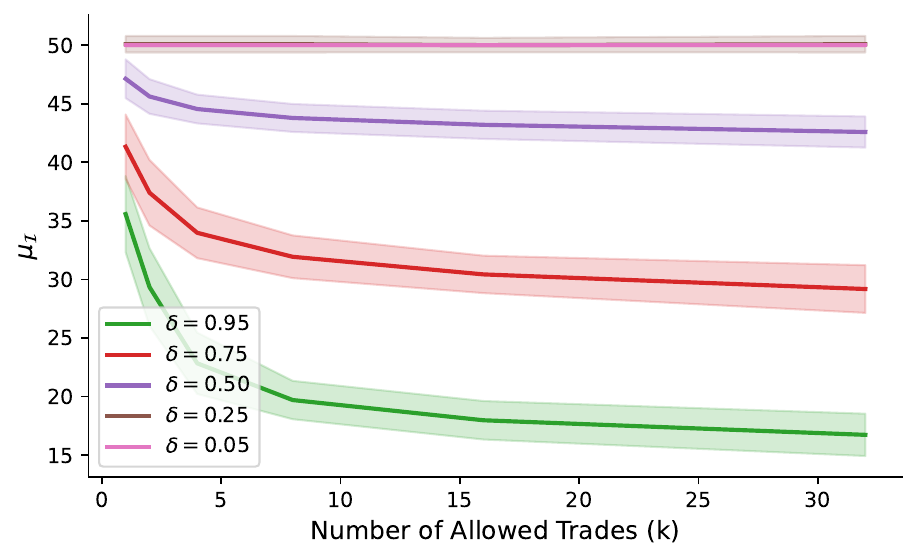}
        \caption{Effect of $k$ on $\mu_{I}$ under different $\delta$}
    \label{fig:rq3_dispersion_muI}
    \end{subfigure}
    \hfill
    \begin{subfigure}{0.95\linewidth}
        \includegraphics[width=\linewidth]{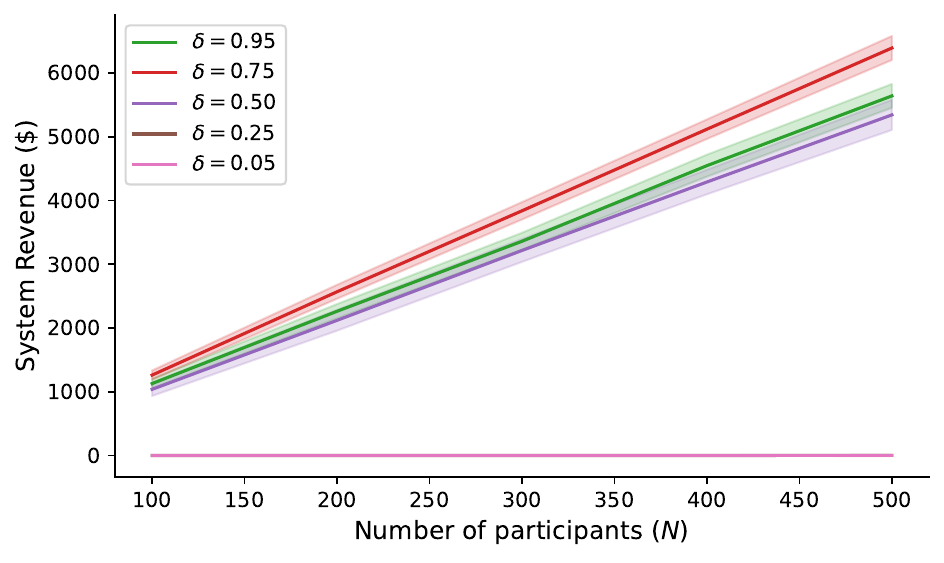}
    \caption{Effect of $N$ on revenue under different $\delta$}
    \label{fig:rq3_dispersion_profit}
    \end{subfigure}
    \caption{Impact of price dispersion ($\delta$) on outcomes ($\gamma = 0.4$). As $k$ increases, more consumers can access the lower price (a). Note that $\mu_I$ without trading is $\approx \$50$.
    If dispersion is too low for a given $\gamma$, trading will decrease (b). Assuming a financially viable $\gamma$, system revenue increases with $N$, and $\delta = 0.75$ maximizes revenue. At $\delta > 0.75$ nearly all consumers have incentive to trade; increasing $\delta$ lowers the Nash bargaining solution and thus system revenue. \textbf{Increasing price dispersion lowers mean cost paid ($\mu_I$) and increases system revenue to a point.}}
    \label{fig:rq3_dispersion}
\end{figure}

In \textbf{RQ 3} we ask: How does revenue to exchange system $\mathcal{S}$ change due to $\gamma$ (the cut taken by the system), $N$ (the number of consumers), and $\delta$ (the dispersion of pricing algorithm $\mathcal{A}$)? 
For exchange system $\mathcal{S}$ to be economically viable, 1) the system must be able to recover operation costs and 2) consumers must have an incentive to participate even when the system takes a cut. 
We compare decentralized methods to determine which earns the system the most revenue.
In \Cref{fig:rq3} we show, for different $\gamma$ values, similar system revenues from each approach. 
As $\gamma$ increases, revenue increases until some maximum (approximately $\gamma = 0.8$), where consumers refuse to trade, and revenue decreases. At this point, the system earns $\approx \$2200$ (see the peak in \Cref{fig:rq3}). Given that the original seller would have made $\$5000$ when consumers were not trading (with an average good price of \$50 for 100 consumers), this represents a substantial cut of revenue being redirected to the exchange.    
In ~\Cref{fig:rq3} we show system revenue \textit{and} consumer intermediary profits specifically for $\mu_I^D$. 
For sufficiently high $\gamma$, system revenue decreases as the system takes more and consumers find it unprofitable to trade. Intermediary consumers consistently lose profits as $\gamma$ increases.

To study the impact of dispersion $\delta$ and number of consumers $N$ on revenue, we again focus on $\mu_{I}^D$. 
We examine the importance of dispersion ($\delta$) in \Cref{fig:rq3_dispersion}, where we run our simulation with five different pricing algorithms: $\mathcal{A}_{0.05}$,$ \mathcal{A}_{0.25}$, $\mathcal{A}_{0.50}$, $\mathcal{A}_{0.75}$, $\mathcal{A}_{0.95}$. 
We construct our dispersion models to hold means constant---this allows for direct comparison of dispersion levels, which can dramatically impact the prices that individuals pay as well as the sustainability of the system. 
In \Cref{fig:rq3_dispersion_muI} we show that under system $\mathcal{S}$, higher dispersion models result in lower average prices for individuals. For high values of $k$, the highest dispersion model achieves an average net price of $\$16$ versus $\$50$ in the lowest dispersion case---a near $66\%$ reduction. This is in contrast to the medium dispersion case ($\delta  = 0.5$) with  a reduction of $\approx 15\%$ and the low dispersion case ($\delta = 0.05$) with no reduction.

~\Cref{fig:rq3_dispersion_profit} shows that this system earns more under higher dispersion settings; 
if dispersion is too low, trading with any fees is not viable. 
We closely examine the high dispersion scenarios: the pricing algorithm with $\delta = 0.75$ narrowly earns more than $\delta = 0.95$ (within variance bounds). 
When dispersion is sufficiently high, consumers are willing to trade. If dispersion increases (from $\delta = 0.75$ to $0.95$), 
it results in 
a decrease in the mean Nash bargaining solution for price, so the system earns slightly less. Nevertheless, the revenues are comparable. 

We can also use the dispersion $\delta$ of the market to determine whether one should invest in this type of system. 
We see in low dispersion settings ($\delta = 0.05$ or $\delta = 0.25$) that no $N$ value would be able to sustain this system; for $\delta = 0.50$ to $\delta = 0.95$ the revenue scales linearly up to $N=500$. To justify developing such a system, market prices need to exhibit sufficiently high dispersion---else there is no revenue to be made regardless of $N$. Whether this system $\mathcal{S}$ is profitable to build depends on the cost structure of the implementation: the fixed cost as well as cost that scales with $N$. 
Our results suggest that more dispersed personalized pricing allows the system $\mathcal{S}$ to better help users achieve fairer outcomes while earning good revenue. This provides an opportunity for fair pricing as well as keeping sellers accountable when employing extreme personalization.

\section{Discussion}
\begin{figure}[h]
    \centering
    \begin{subfigure}{0.95\linewidth}
        \centering
        \includegraphics[width=\linewidth]{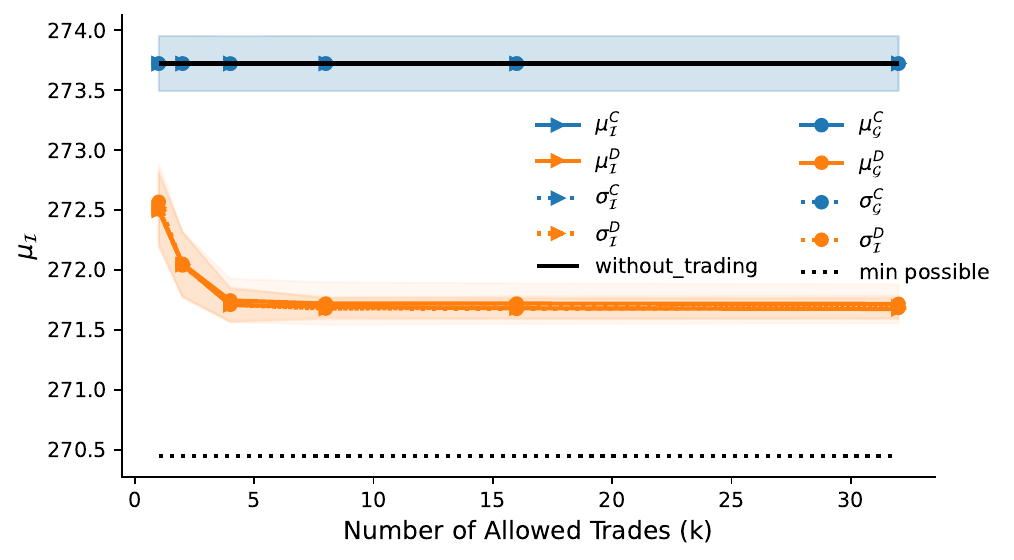}
        \caption{Welfare improvement for flights, ($\gamma = 0.005$)}
        \label{fig:dis_flight_welfare}
    \end{subfigure}
    \begin{subfigure}{0.95\linewidth}
        \centering
        \includegraphics[width=\linewidth]{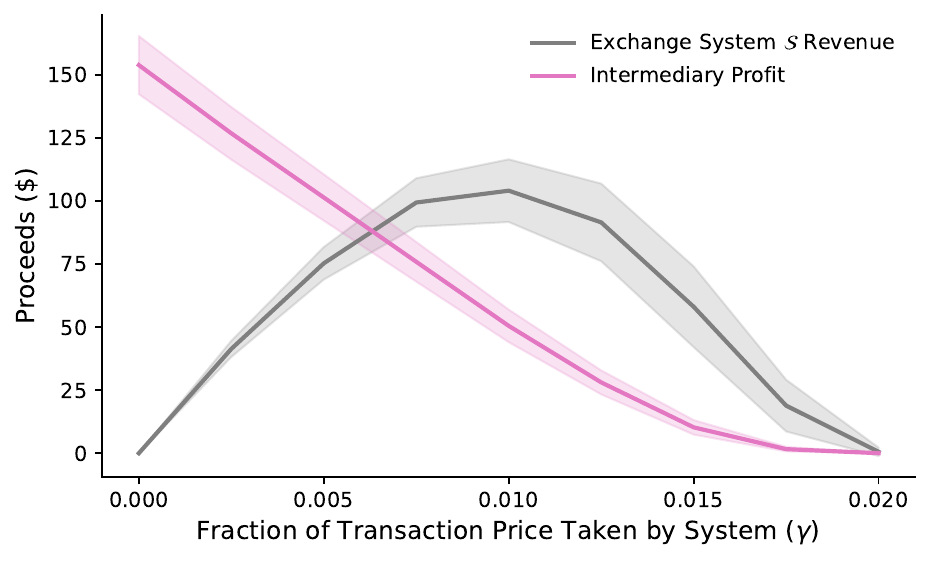}
        \caption{$\gamma$ vs system proceeds}
        \label{fig:dis_flight_profit}
    \end{subfigure}
    \caption{Exchange system on an empirical pricing distribution. In (a) we see again that decentralized methods are able to reduce the gap between the best possible price and the prices paid by $62\%$. In (b) the cut taken by the system $\gamma$ must be low; at $\gamma = 0.01$ the system still earns $\$100$.}
    \label{fig:dis_flight}
\end{figure}

\textbf{Analysis of an empirical pricing distribution}:
Here we show our system $\mathcal{S}$ on a real (rather than simulated) pricing distribution. 
We use prices as modeled in prior work \cite{browsing23}.
Using this pricing distribution (which includes nine non-overlapping groups), we run a simulation with $N=100$ consumers. \Cref{fig:dis_flight} shows improvement even when the range in prices is small.
Specifically, the range is $\$6.15$, and the maximum price is $\$275.82$; this gives a dispersion of $\approx 0.02$ after normalizing prices to $(0, 1]$. Compared to the simulated example, the minimum price achievable is not $0$ but rather $\approx 270$; hence this is the lowest possible expected price.
Without trading, the gap between the average price and best price was $\$3.23$. Trading using the $\mu_{I}^D$ methodology results in $\$1.22$, a $62\%$ reduction of the gap (\Cref{fig:dis_flight_welfare}) while reducing the sticker price by $\approx 0.75\%$. 
We set $\gamma$ values to be much smaller than in the simulated examples due to the higher transaction prices. Even with low $\gamma$ values, revenues are reasonable, peaking at $\gamma = 0.01$ (\Cref{fig:dis_flight_profit}) and about $\$100$ in revenue for $N = 100$. 
These results indicate the feasibility of implementing our system in a market such as this one.

\textbf{What is fair?}:
We consider various definitions of fairness in this work, outlined in \Cref{tab:fairness}. 
Ideally, one would like to maximize all definitions simultaneously for the most ``fair'' result.
In particular, the perfect solution would involve every member of every group simultaneously achieving the lowest price, minimizing the variation and average price paid across individuals and groups.
Our results suggest simultaneously achieving low prices and low variation in prices across individuals or groups is difficult. 
In particular, there is an explicit trade-off between mean and variance. In order to reduce the net price, an intermediary with a low price must be 
incentivized to engage in trades by earning money. However, at this scale, an intermediary can benefit from multiple trades, while a buyer can only benefit from a single trade---this can increase variance. We make no claim regarding which is ideal; it is context dependent.

\textbf{A more informed system:} 
In our simulations, we assume that the system $\mathcal{S}$ has minimal information: it is aware of the prices that the consumers receive, but not each individual consumer's disutility ($\epsilon_v$) per trade. This limits the ability of the system to effectively match consumers, especially in the centralized case, where the optimal matching requires one of the consumers to transact with minimal personal benefit. Thus, such transactions are not executed.
If the exchange system were aware of each consumer's disutility, it could incorporate that information in determining the optimal prices and result in more successful, fairness-improving transactions. 
This, however, requires collecting information from consumers that they may not be entirely aware of. Though we model consumer $v$'s disutility with a distribution $\varepsilon_v$, consumer $v$ ``draws'' from this distribution for each transaction. This draw represents the current state of consumer $v$ and how long they are willing to spend on this transaction, which is information they may not be able to accurately convey. While having more information could boost the effectiveness of the centralized matching, the details regarding collecting it are better left for future studies. 
 
\textbf{Closeness of $\mu_I$ and $\mu_{\mathcal{G}}$}: In our implementation, $\mu_I$ and $\mu_{\mathcal{G}}$ values track very closely.
This is because an average over individuals weights each individual equally, while an average over groups weights individuals in smaller groups more heavily.
Given our assumption that group sizes are roughly equal, $\mu_{I}$ and $\mu_{G}$ are algebraically quite close; if group sizes were identical, $\mu_{I}$ and $\mu_{G}$ would be as well. Using different distribution of group sizes can lead to different outcomes. We show an example in \Cref{sec:group_distributions} where in highly skewed cases, a small group getting the best price can receive a far more disproportionate benefit. Future work can look at different and more realistic distributions of pricing buckets.

\textbf{Interaction between the exchange system and market}:
In this work, we assume that the exchange system and market are independent; in particular, that the market's pricing algorithm does not respond to consumer trading. We believe this is a reasonable assumption; if a very small group of individuals is participating in this system, the consumers' behavior may go unnoticed. We simulate this a single-shot game, with no opportunity for the seller to respond. We recognize that if the behavior is detected, the pricing system could react, possibly raising prices overall and resulting in all users losing access to cheaper pricing \cite{kosmopoulou2016customer}.

If we allow for a repeated game, there are several ways a marketplace might respond: 1) increasing technical barriers for the consumer that block trades, 2) increasing personalization, 3) decreasing personalization but increasing prices overall. While we cannot predict which of these might occur, the modeling of our system is robust to all of these changes–increased technical barriers can be modeled as additional disutility to each consumer from participating in each transaction; increased personalization would actually increase the fairness of outcomes; decreased personalization, while decreasing fairness, can be easily modeled with a change to pricing algorithm $\mathcal{A}$.

Another issue that may arise between the system and market is that owners of the exchange system could forgo fairness and accountability in exchange for higher revenue. If the system and  market collude, the market could produce prices that, when passed through the colluding exchange system $\mathcal{S}$, result in worse outcomes for consumers, but higher revenue for the system and market. Considering our finding that higher dispersion can increase revenue and decrease net cost to consumers, existence of such a strategy is possible.
Thus, the assumption that the system is independent from the market and has different goals is key.

\section{Future Work and Limitations}
\textbf{Scale of transactions:}
In this work we discuss the sale of one good $g$.
We assume consumers $V$ simultaneously browse the market $\mathcal{M}$ for good $g$, 
so the system $\mathcal{S}$ matches these consumers. This assumption is important so all consumers in $V$ can acquire good $g$ in a timely manner. It is possible, however, that consumers are browsing the same marketplace for multiple goods. In this case, multiple matchings and transactions would occur in parallel, increasing profit but also the cost of the system $\mathcal{S}$ to maintain itself. If our system were to be deployed on a real market, the time scale and number of goods would need to be considered in the matching. We leave the multi-good matching problem for future work.

\textbf{Market structure:}
Our simulations assume a particular market structure, where a single good is available with enough supply to satisfy all demand. We also assume that each consumer \textit{must} buy the good. However, one rationale for personalized pricing is that some users are less willing to pay for a good than others \cite{NBERw23775}. Incorporating this requires a more detailed utility function for good purchase, which we defer to future work. 
We also assume that the resource constraint per user is the same; in practice this may not be true. 

\textbf{System participation:}
Here we allowed consumers to participate for free but pay a fee per transaction. Another potential structure is to charge a flat fee to participate, but then allow consumers to exchange freely. However, this structure requires convincing users to join the system with an upfront fee. Investigating and comparing these exchange system structures is an avenue for future work. 

\textbf{Human studies:}
In this work we make simplifying assumptions regarding consumers' actions, \textit{e.g.,} the form of their utility functions, willingness to participate, demographic attributes, and resource constraints. To test these assumptions, future human studies would be ideal. This would allow us to test our design principles on users who may not respond as we originally modeled them. Other design principles could be tested as well (\textit{e.g.,} wording of messaging to consumers) to incentivize trading.

\section{Conclusion}

In this paper we introduced a fairness-centered exchange system that takes advantage of personalized pricing to improve fairness. We simulated the effect of price dispersion and explored two different transaction price-setting procedures paired with four fairness targets to answer \textbf{RQ 1} and \textbf{RQ 2}. These modeling choices set the transaction prices from which the system and consumers can profit. 
To answer \textbf{RQ 3}, we showed that our system's revenue is higher when prices are more dispersed, and that consumers are able to achieve lower prices---up to $66\%$ improvement for the mean net cost to individuals and $69\%$ for the mean net cost to a group.
We demonstrated that a decentralized negotiation approach is better able to achieve most notions of fairness compared to a centralized approach.
While a designer could choose to focus on either mean or standard deviation as fairness targets, from a financial sustainability perspective, minimizing mean net cost paid by an individual earned more than other targets. Our approach is a consumer-driven solution to personalized pricing that does not rely on fair pricing by the seller, or marketplace regulations. Further directions include considering multiple goods---requiring computing multiple matchings in parallel, and changing the market structure such that all consumers are not required to buy the good.
One might also consider varied financial structures to support the exchange system, or conducting human studies to better test the assumptions we made in this work.
Our results provide theoretical evidence that such a system could improve fairness for consumers while sustaining itself financially and holding sellers accountable for extreme personalized pricing.

\bibliographystyle{ACM-Reference-Format} 
\bibliography{ref}

\appendix

\newpage
\clearpage
\section{Notation}\label{sec:notation}
The following table serves as a reference for notation used throughout the paper.

\begin{table}[h]
    \caption{Summary of Notation}
    \label{tab:notation}
    \begin{tabularx}{\linewidth}{@{}r X @{}}
    \textsc{Notation} & \textsc{Description}\\ 
    \midrule
        $G = (V,E)$ & the set of agents $V$ connected by edges $E$ \\
        $\mathcal{M}$ & the marketplace \\
        $g$ & the one type of good being offered in $\mathcal{M}$\\
        $\mathcal{S}$ & the exchange system \\
        $\mathcal{A}$ & the pricing algorithm \\
        $\mathcal{P}$ & the agent-matching process \\
        $\mathcal{J}$ & the set of agent interactions defined by the matching process \\
        $\mathcal{F}$ & the fairness metric that our process aims to optimize \\
        $r_v$ & the vector of consumer $v$'s properties \\
        $\delta$ & the dispersion of pricing algorithm $\mathcal{A}$ \\
        $p_v$ & the price offered to agent $v$ by algorithm $\mathcal{A}$ \\
        $N$ & the number of agents $V$\\
        $f_v(j_{uv})$ & utility function $f$ for agent $v$ that takes an interaction and price as input \\
        $m$ & the transaction price of an interaction \\
        $k$ & resource constraint for all agents \\
        $\gamma$ & the proportion of transaction price $m$ that the system $\mathcal{S}$ takes
\end{tabularx}
\end{table}

\section{Impact of different group size distributions} \label{sec:group_distributions}

As mentioned in the discussion, in the main body of the paper we study the case when group sizes are roughly equal. As an additional measure, we briefly examined a case where the group sizes are unequal ($N=100, \delta = 0.95, \gamma = 0.4$). Specially, we consider groups sizes sampled from a power law distribution with $\beta =2$. In one instance, the group size is assigned directly proportional to the price (\textit{i.e.,} the largest group has the highest price) while the other instance is reversed (\textit{i.e.,} the largest group has the lowest price). 

\begin{figure*}[h]
    \centering
    \begin{subfigure}{0.45\linewidth}
        \centering
        \includegraphics[width=\linewidth]{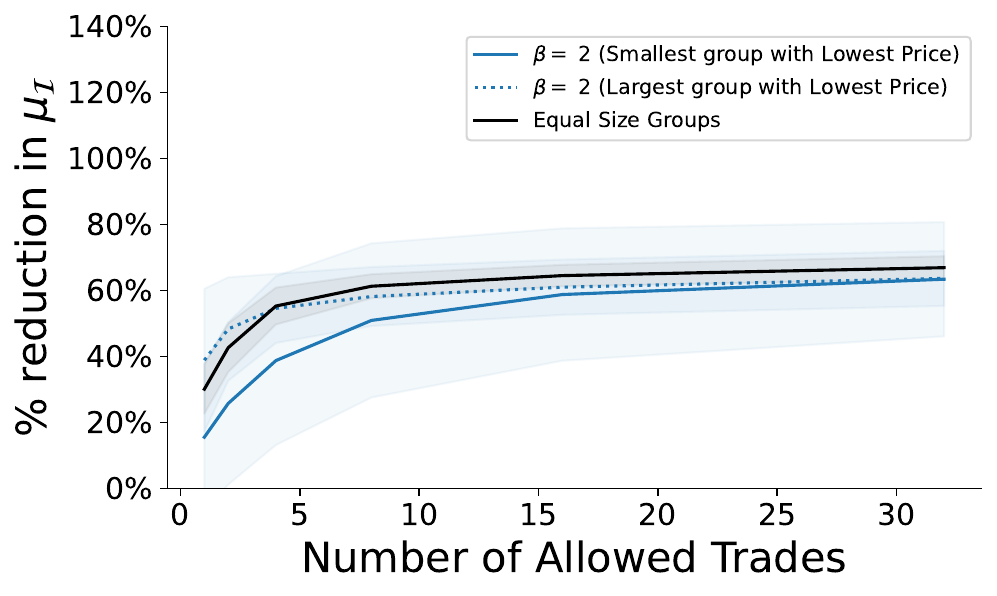}
        \caption{$\mu_{I}$}
        \label{fig:skewed_individual_fairness}
    \end{subfigure}
    \begin{subfigure}{0.45\linewidth}
        \centering
        \includegraphics[width=\linewidth]{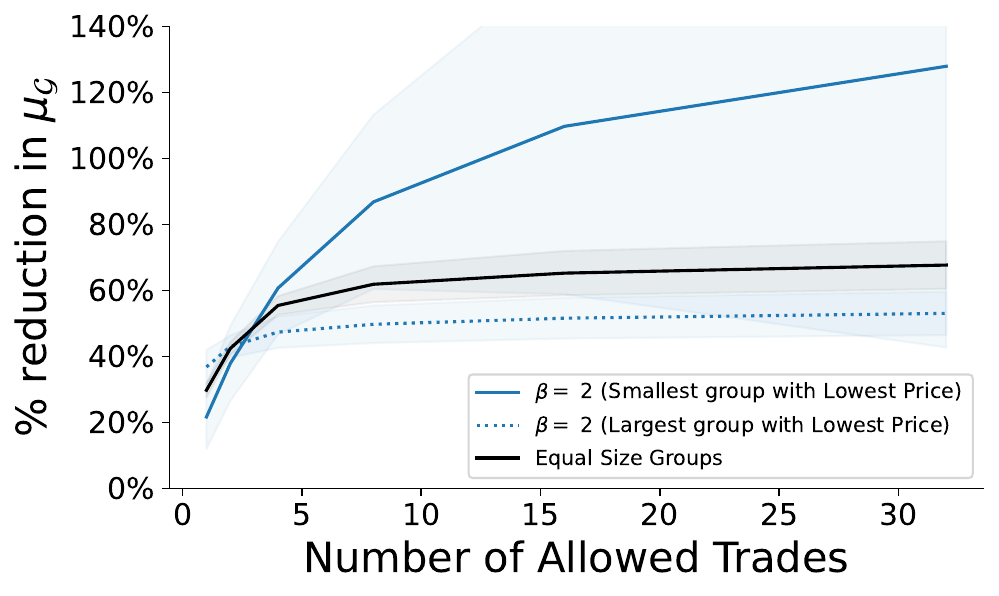}
        \caption{$\mu_{G}$}
        \label{fig:skewed_group_fairness}
    \end{subfigure}
    \caption{Impact of group size distribution on the reduction in as $k$ (number of allowed trades) increases. We optimize for $\mu_{I}$ and $\mu_{G}$ in the decentralized case, and set $\gamma = 0.4, \delta = 0.95$. Three group size distributions are shown: one where all groups are equally sized, one where the largest group is offered the lowest price, and where the largest group is offered the highest price. Here the percentage reduction in $\mu_{I}$, $\mu_{G}$ is shown compared to a no-trading situation. }
    \label{fig:skewed_dis_flight}
\end{figure*}

 In \Cref{fig:skewed_individual_fairness}, we see that in $\mu_{I}$, while the percentage reduction varies slightly with a lower number of trades $k$, as $k$ increases, $\mu_{I}$, is reduced by a similar amount for the skewed and equally sized groups. 
 
In contrast, in \Cref{fig:skewed_group_fairness} we see that $\mu_{G}$, the mean net cost paid by groups, is reduced much further in the scenario where the small group is offered the lowest price. As $k$ gets larger, we see that the percentage reduction is above $100\%$, because this small group earns enough outsized benefit to drive the $\mu_{G}$ negative. While all groups can benefit in this scenario, the small, privilege group has the opportunity to earn much more and drive down the average group price more substantially. 

\section{Pricing algorithm details} \label{sec:pricing}
As described in Algorithm 1, we construct a family of pricing distributions $\mathcal{A}_{\delta}$ parameterized by $\delta$, where $\delta$ represents the $2.25\sigma$ range of possible prices. We consider $\delta = \{0.05, 0.25, 0.5, 0.75, 0.95\}$ with $|\mathcal{G}| = 5$. For a given pricing algorithm, we have $\mu_1, \mu_2, ... \mu_5$ for each group and a fixed $\sigma$ for all groups. For each group member in group $g \in \mathcal{G}$, we sample a price from $N(\mu_g, \sigma)$. Here we detail for all five pricing algorithms $A_{\delta}$, the respective group's mean as well as the $\sigma$ used. By construction, the price distributions produced by these algorithms have the same mean ($\$50$), which allows us to specifically focus on understanding the role of dispersion within our proposed system.

\begin{table}[!h]
\begin{center}
\caption{Parameters that go into $\mathcal{A}_{\delta}$ for the respective pricing algorithm}
\begin{tabular}{lrrrrrr}
$\delta$ &   $\mu_{1}$ &  $\mu_{2}$ &  $\mu_{3}$ &  $\mu_{4}$ &  $\mu_{5}$ &  $\sigma$  \\
\midrule
$0.95$ & $10$ & $30$ & $50$ & $70$ & $90$ & $\frac{30}{9}$\\
$0.75$ & $20$ & $35$ & $50$ & $65$ & $80$ & $\frac{30}{9}$\\
$0.5$ & $30$ & $40$ & $50$ & $60$ & $70$ & $\frac{20}{9}$\\
$0.25$ & $40$ & $45$ & $50$ & $55$ & $60$ & $\frac{10}{9}$\\
$0.05$ & $50$ & $50$ & $50$ & $50$ & $50$ & $\frac{10}{9}$\
\end{tabular}
\end{center}
\label{tab:pricing_algo}
\end{table}

\section{Flight Pricing Simulation}
\subsection{Flight Price Model}{\label{sec:appex_flight}}
We use a pricing model from prior work to produce our empirical pricing distribution~\cite{browsing23}. In this market, the average price was $\$270.45$. The work presents pricing models from various sellers in the market, we use the model presented for ``Third Party 1''. In this work, the authors present the offsets from the base price of the flight ticket : $[\$4.55, \$1.46, \$5.29, \$3.55, \$6.15, \$2.91, \$2.36, \$5.05]$. 

\begin{table}
\begin{center}
\caption{Prices for $\mathcal{A}_{\text{flight}}$. No variance is used for this algorithm. }
\begin{tabular}{c|c}
Group  & Price \\
\hline
$\mu_1$ & \$270.45 \\
$\mu_2$ & \$271.91 \\
$\mu_3$ & \$272.46 \\
$\mu_{4}$ & \$273.01 \\
$\mu_{5}$ & \$274.21 \\
$\mu_{6}$ & \$275.42 \\
$\mu_7$ & \$275.82\\
$\mu_8$ & \$276.20\\
$\mu_9$ & \$276.60

\end{tabular}
\end{center}

\label{tab:pricing_algo_flight}
\end{table}

From these mean values we created nine non-overlapping groups of consumers, where the price for each group is the offset from the base price. We directly assign the price from one of the nine mean values listed purely based on the group.

\subsection{Consumer Utilities}
The utility structure of the consumers is the same as described in the main body of the paper. 
For the individual disutility, each consumer $u$ is assigned a truncated Normal distribution $\mathcal{E}_u$ with varying means (drawn from $U[0, 1]$) but the same standard deviation ($0.5$) where the numbers are chosen with respect to the prices from the flight price model.

\end{document}